\documentclass[preprint]{elsarticle}

\usepackage{lineno,hyperref}
\usepackage{graphicx,amsmath,amssymb,amsthm,bm}
\usepackage{algorithm,algorithmic}
\modulolinenumbers[5]

\journal{}

\newtheorem{definition}{Definition}
\newtheorem{proposition}{Proposition}

\newtheorem{theorem}{Theorem}

\begin{document}

\begin{frontmatter}

\title{Unbeatable imitation of a friend}

\author{Masahiko Ueda\corref{mycorrespondingauthor}}
\ead{ueda@tmu.ac.jp}
\address{Department of Physics, Tokyo Metropolitan University, Tokyo 192-0397, Japan}

\begin{abstract}
Imitation sometimes achieves success in multi-agent situations even though it is very simple.
In repeated games, success of imitation has been characterized by unbeatability against other agents.
Previous studies specified conditions under which imitation is unbeatable against co-players in repeated symmetric games, and clarified that the existence of unbeatable imitation is strongly related to the existence of payoff-controlling strategies, called zero-determinant strategies.
However, these results were on the situations in which one repeated game is played by one group of players.
It was pointed out that imitation of other players in the same group and imitation of other players in the same role in other groups generally result in different outcomes.
Here, we investigate the existence condition of unbeatable imitation in the latter situations, described by two concurrent repeated games played by two different groups of players.
We find that the existence condition of unbeatable imitation is stronger than the existence condition of unbeatable zero-determinant strategies, whereas both are very limited.
Furthermore, we find that adding exploration can make an imitation strategy unbeatable even in games where imitation without exploration can be beaten.
Our findings suggest a strong relation between unbeatable imitation and payoff control even in the two concurrent repeated game situations.
\end{abstract}


\end{frontmatter}


\section{Introduction}
\label{sec:intro}
Imitation is simple yet successful behavior in multi-agent systems.
In the context of evolution of cooperation in social dilemmas, it has been widely known that the Tit-for-Tat (TFT) strategy, which imitates the previous action of the opponent, won a computer tournament \cite{AxeHam1981}.
Although TFT is vulnerable to noise compared to the Win-Stay Lose-Shift strategy \cite{NowSig1993}, which is one of the learning-based strategies \cite{MacFla2002,MasNak2011}, an imitation-based strategy can be easily implemented because it does not use concrete values of payoffs.
In oligopoly markets, it was pointed out that perfect competition can be realized by imitation, rather than by rational decision-making \cite{Veg1997}.
In some multi-armed bandits where agents can use limited and uncertain information, the optimal learning rule is imitation of other agents \cite{Sch1998}.
Even in the same oligopoly games, the results are different depending on who an agent imitates \cite{AHO2007}.
Furthermore, it has been known that, if all agents become imitators in decision-making, they cannot adapt to new environments \cite{RBCet2010}.

One concept which characterizes the success of imitation in game theory is that it is ``unbeatable'' against an opponent \cite{DOS2012b,DOS2014}.
A strategy is called unbeatable if it always obtains a payoff not less than that of the opponent.
In evolutionary game theory, an unbeatable strategy can invade any other strategies by neutral drift, because it obtains a payoff not less than that of the opponent.
In repeated two-player symmetric games, the conditions in which imitation strategies are unbeatable were specified.
The Imitate-If-Better (IIB) strategy, which imitates the previous action of the opponent when it was beaten, is unbeatable if and only if the game does not contain any generalized rock-paper-scissors cycles \cite{DOS2012b}.
TFT is unbeatable if and only if the game is a potential game \cite{DOS2014}.
In the repeated prisoner's dilemma game, IIB corresponds to the Grim Trigger strategy \cite{Fri1971}, and both IIB and TFT are unbeatable.
In addition, they are cooperative Nash equilibrium strategies.
In repeated multi-player totally-symmetric games, a class of games where IIB is unbeatable against all opponents was recently specified, which can be regarded as a multi-player version of games with no generalized rock-paper-scissors cycles \cite{Ued2023}.

In parallel to the discovery of unbeatable imitation strategies, an important class of strategies, called zero-determinant (ZD) strategies, was found in repeated games \cite{PreDys2012}.
ZD strategies unilaterally enforce linear relationships between payoffs of players in order to control payoffs.
ZD strategies have attracted much attention both in evolutionary game theory and in multi-agent systems \cite{HRZ2015,HTS2015,Aki2016,McAHau2016,MamIch2020,LiuWu2022,Ued2022c}, since they do not assume rationality of other players.
Particularly, it was pointed out that TFT is a fair ZD strategy, which unilaterally equalizes the payoffs of two players, in the repeated prisoner's dilemma game \cite{PreDys2012}.
This result suggests the possibility that an unbeatable property of imitation may generally be realized by a payoff control ability of a ZD strategy.
In fact, it was proved that, in repeated general two-player symmetric games, TFT is unbeatable if and only if TFT is a fair ZD strategy \cite{Ued2022}.
It was also shown that, in repeated two-player symmetric games, if IIB is unbeatable, then a fair ZD strategy also exists \cite{Ued2022b}.
In addition, in the above class of repeated multi-player totally-symmetric games where IIB is unbeatable, an unbeatable ZD strategy also exists \cite{Ued2023}.
These results suggest that the existence of unbeatable imitation strategies is strongly related to the existence of unbeatable ZD strategies, and that probably the latter is realized in some weaker conditions than the former.

\begin{figure}[tbp]
\begin{center}
\includegraphics[clip, width=10.0cm]{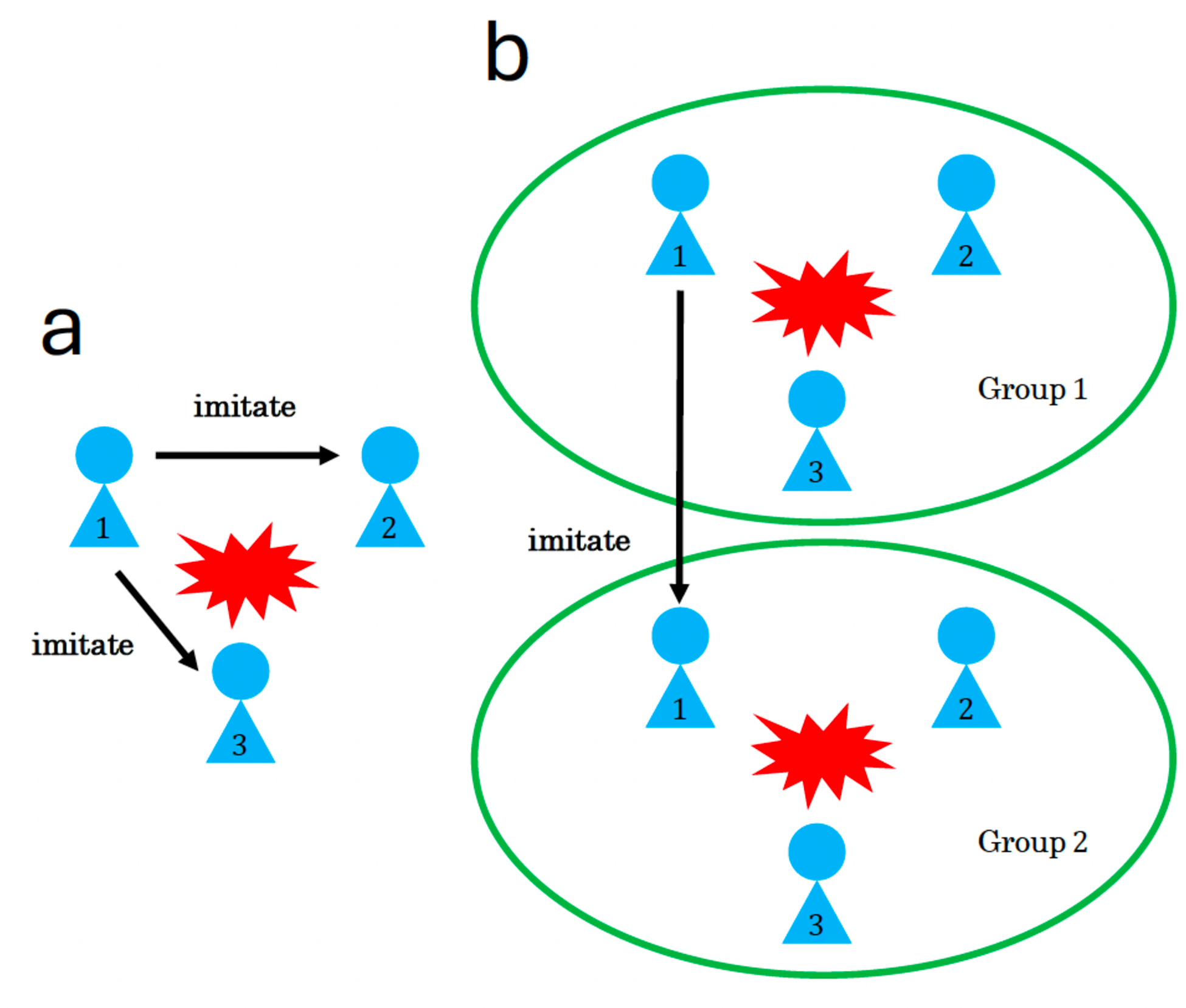}
\end{center}
\caption{Schematic pictures of two situations of imitation. (a) ``Imitation of opponents'' (imitation of other players in the same group). (b) ``Imitation of friends'' (imitation of other players in the same role in other groups).}
\label{fig:imitation}
\end{figure}

It should be noted that all above results about unbeatable imitation and unbeatable ZD strategies are obtained for situations of ``imitation of opponents''.
In \cite{AHO2007}, it was pointed out that imitation of other players in the same group and imitation of other players in the same role in other groups generally result in different outcomes.
Here, we call the former and the latter as ``imitation of opponents'' (Fig. \ref{fig:imitation}a) and ``imitation of friends'' (Fig. \ref{fig:imitation}b), respectively.
In the Cournot oligopoly game, the former leads to the Walras equilibrium, and the latter leads to the Cournot-Nash equilibrium \cite{AHO2007}.
Performance of imitation and ZD strategies in the ``imitation of friends'' situations is largely unknown.

A typical situation of ``imitation of friends'' is as follows.
Two firms 1 and 2 produce both apple juice and orange juice.
The structure of two juice markets is assumed to be almost equivalent.
Both Alice and Bob belong to firm 1.
In firm 1, Alice and Bob determine the production of apple juice and orange juice, respectively.
Alice does not have enough knowledge about the apple juice market, and she wants to cut corners as much as possible while doing the work.
So she came up with naively imitating Bob's work.
But, at the same time, she does not want to lose to him in firm 1.
A question is ``What is the condition that Alice does not lose to Bob by only imitating him?''.
A similar situation is found in social learning by bees \cite{WorPap2005}, where copying of flower choice by bees from a different nest was observed.

In this paper, we investigate the performance of imitation and ZD strategies in the ``imitation of friends'' situations.
Particularly, we specify the condition for the existence of unbeatable imitation and the condition for the existence of a fair ZD strategy in the ``imitation of a friend'' situations.
We show that both are very limited, and specifically, the latter is weaker than the former.
Furthermore, we prove that the unbeatable imitation strategies are realized as unbeatable ZD strategies.
Together with the above results in ``imitation of opponents'' situations, our results suggest a strong relation between the existence of unbeatable imitation strategies and the existence of unbeatable ZD strategies.
We also prove that adding exploration can make an imitation strategy unbeatable even in games where imitation without exploration can be beaten.

The paper is organized as follows.
In Section \ref{sec:preliminaries}, we introduce a model of two concurrent games and several concepts necessary in this paper.
In Section \ref{sec:results}, we provide our main results.
In Section \ref{sec:discussion}, we discuss the implication of the results.
The proofs of the theoretical results are provided in Section \ref{sec:methods}.
In \ref{app:stronger_IIBTFT}, stronger versions of some theoretical results are provided.
In \ref{app:bandit}, performance of imitation in a bandit-like game is numerically investigated.

\section{Preliminaries}
\label{sec:preliminaries}
We consider two identical repeated games played in parallel (Fig. \ref{fig:imitation}b) \cite{AHO2007}.
The stage game is described as $G:=(G_1, G_2)$ with
\begin{align}
 G_\mu &:=\left( \mathcal{N}_\mu, \left\{ A_{(\mu, j)} \right\}_{(\mu, j)\in \mathcal{N}_\mu}, \left\{ s_{(\mu, j)} \right\}_{(\mu, j)\in \mathcal{N}_\mu} \right) \quad (\mu=1, 2).
\end{align}
$\mathcal{N}_\mu := \left\{ (\mu, 1), \cdots, (\mu, N) \right\}$ is the set of players in group $\mu$, and $N\geq 2$ is the number of players in each group.
$A_{(\mu, j)}$ is the set of actions of player $(\mu, j)$, and satisfies $A_{(1, j)}=A_{(2, j)}=A_{j}$ for all $j\in \{ 1, \cdots, N \}$.
$s_{(\mu, j)}: \prod_{j=1}^N A_j \rightarrow \mathbb{R}$ is the payoff function of player $(\mu, j)$, and satisfies $s_{(1, j)}=s_{(2, j)}=s_{j}$ for all $j\in \{ 1, \cdots, N \}$.
We write an action profile as $\bm{a}:=\left( \bm{a}_1, \bm{a}_2 \right)$ with $\bm{a}_\mu := \left( a_{(\mu, 1)}, \cdots, a_{(\mu, N)} \right)$ $(\mu=1, 2)$.
For convenience, we introduce the notations $\mathcal{A}:=\prod_{j=1}^N A_j$, $A_{-j}:=\prod_{k\neq j} A_k$, and $a_{(\mu, -j)}:=\left( a_{(\mu, 1)}, \cdots, a_{(\mu, j-1)}, a_{(\mu, j+1)}, \cdots, a_{(\mu, N)} \right) \in A_{-j}$ for all $\mu$ and all $j$.
In this paper, we assume that $A_j$ is a finite set for all $j$.

In the repeated version of $G$, the action profile at the $t$-th round is described as $\bm{a}^{(t)}$.
The behavior strategy of player $(\mu, j)$ is conditional probability distributions $T_{(\mu, j)}^{(t)}$ on $A_j$ given the histories $\left\{ \bm{a}^{(t^\prime)} \right\}_{t^\prime=1}^{t-1}$.
The payoff of player $(\mu, j)$ in the repeated game is given by
\begin{align}
 \mathcal{S}_{(\mu, j)} &:= \lim_{T\rightarrow \infty} \frac{1}{T} \sum_{t=1}^T \mathbb{E}\left[ s_j\left( \bm{a}_\mu^{(t)} \right) \right],
 \label{eq:payoff_repeated}
\end{align}
where $\mathbb{E}\left[ \cdots \right]$ describes an expected value with respect to the behavior strategies of all players.
In other words, we assume that there is no discounting.
If such a limit does not exist, we extract a convergent subsequence of the time-averaged distribution of action profiles \cite{Aki2016}.

This model is similar to multichannel games with two channels \cite{DHNH2020}.
However, in multichannel games, players $(1, j)$ and $(2, j)$ can be regarded as one player, and the payoff of player $j$ is defined by $\mathcal{S}_{(1, j)}+\mathcal{S}_{(2, j)}$.
In addition, in multichannel games, the games $G_1$ and $G_2$ can be different.
In contrast, in our model, players $(1, j)$ and $(2, j)$ are different players in the same role, and the structure of two games $G_1$ and $G_2$ must be identical, which is a minimal setup of imitation.
Therefore, while the models are similar to each other, the situations considered are quite different.
In addition, we want to remark that most of other multi-agent frameworks correspond to the ``imitation of opponents'' situations (Fig. \ref{fig:imitation}a).

Next, we introduce several concepts used in this paper.
First, we introduce the concept of memory-one strategies.
A behavior strategy $\left\{ T_{(1, j)}^{(t)} \right\}_{t=1}^\infty$ of player $(\mu, j)$ is a \emph{memory-one} strategy if it satisfies
\begin{align}
 T_{(\mu, j)}^{(t)} \left( a_{(\mu, j)}^{(t)} | \left\{ \bm{a}^{(t^\prime)} \right\}_{t^\prime=1}^{t-1} \right) &= T_{(\mu, j)} \left( a_{(\mu, j)}^{(t)} | \bm{a}^{(t-1)} \right)
\end{align}
for all $t\geq 2$, all $a_{(\mu, j)}^{(t)} \in A_j$, and all $\left\{ \bm{a}^{(t^\prime)} \right\}_{t^\prime=1}^{t-1} \in \left( \mathcal{A} \times \mathcal{A} \right)^{t-1}$.
Therefore, a memory-one strategy only remembers the last action profile and chooses an action in each round.

Next, we introduce the concept of unbeatable strategies \cite{DOS2012b,DOS2014} in the ``imitation of a friend'' situations.
Below, when the focal player is $(1, j)$, we call player $(2, j)$ ``a friend''.
\begin{definition}
\label{def:unbeatable}
A memory-one strategy $T_{(1, j)}$ of player $(1, j)$ is \emph{unbeatable} against player $(2, j)$ if the inequality
\begin{align}
 \mathcal{S}_{(1, j)} &\geq \mathcal{S}_{(2, j)}
\end{align}
holds regardless of behavior strategies of the other players and for arbitrary initial actions of player $(1, j)$.
\end{definition}
In other words, if a memory-one strategy is unbeatable, it is guaranteed to obtain the larger payoff than that of a friend.
Unbeatable strategies are similar to ``rival'' (or, ``competitive'') strategies \cite{HTS2015}, but are not the same, because of the arbitrary initial action condition.
Implicitly, it is assumed that player $(1, j)$ has motivation to beat the friend $(2, j)$ as a player in the same role, as in the example in Section \ref{sec:intro}.

We focus on three memory-one strategies; two imitation strategies and one payoff-control strategy.
Below we write the Kronecker delta by $\delta_{a, a^\prime}$.
\begin{definition}
\label{def:TFT}
A memory-one strategy $T_{(1, j)}$ of player $(1, j)$ is the \emph{Tit-for-Tat (TFT)} strategy against player $(2, j)$ if it is described as
\begin{align}
 T_{(1, j)} \left( a_{(1, j)} | \bm{a}^\prime \right) &= \delta_{a_{(1, j)}, a_{(2, j)}^\prime}.
 \label{eq:TFT}
\end{align}
\end{definition}

\begin{definition}
\label{def:IIB}
A memory-one strategy $T_{(1, j)}$ of player $(1, j)$ is the \emph{Imitate-If-Better (IIB)} strategy against player $(2, j)$ if it is described as
\begin{align}
 T_{(1, j)} \left( a_{(1, j)} | \bm{a}^\prime \right) &= \delta_{a_{(1, j)}, a_{(2, j)}^\prime} \mathbb{I} \left( s_j\left( \bm{a}_1^\prime \right) < s_j\left( \bm{a}_2^\prime \right) \right) + \delta_{a_{(1, j)}, a_{(1, j)}^\prime} \mathbb{I} \left( s_j\left( \bm{a}_1^\prime \right) \geq s_j\left( \bm{a}_2^\prime \right) \right),
 \label{eq:IIB}
\end{align}
where $\mathbb{I}( \cdots )$ is the indicator function.
\end{definition}
In other words, TFT of player $(1, j)$ unconditionally imitates the previous action of player $(2, j)$, and IIB of player $(1, j)$ imitates the previous action of player $(2, j)$ if it was beaten by player $(2, j)$ in the previous round.
It should be noted that, although TFT and IIB were competitive strategies originally introduced in the ``imitation of an opponent'' situation, we reuse the terms in the ``imitation of a friend'' situation.
That is, even though a friend $(2, j)$ is nothing but a ``reference player'' (not a direct opponent) of the focal player $(1, j)$, we call the strategy (\ref{eq:TFT}) TFT.
In addition, we remark that TFT and IIB are defined with an arbitrary initial action similarly to Refs. \cite{DOS2012b,DOS2014}, in contrast to standard definitions of TFT and IIB.

\begin{definition}[\cite{PreDys2012,McAHau2016,UedTan2020}]
\label{def:fairZD}
A memory-one strategy $T_{(1, j)}$ of player $(1, j)$ is a \emph{fair zero-determinant (ZD)} strategy against player $(2, j)$ if it can be described in the form
\begin{align}
 \sum_{a_{(1, j)}} c_{a_{(1, j)}} T_{(1, j)} \left( a_{(1, j)} | \bm{a}^\prime \right) - c_{a_{(1, j)}^\prime} &= s_j\left( \bm{a}_1^\prime \right) - s_j\left( \bm{a}_2^\prime \right) \quad \left( \forall \bm{a}^\prime \in \mathcal{A} \times \mathcal{A} \right)
 \label{eq:fairZDS}
\end{align}
with some coefficients $\left\{ c_{a_{(1, j)}}  \right\}$ and both sides are not identically zero.
\end{definition}

Surprisingly, a fair ZD strategy, if exists, unilaterally enforces $\mathcal{S}_{(1, j)}=\mathcal{S}_{(2, j)}$, that is, the payoff of the player using a fair ZD strategy is equal to that of a friend for arbitrary strategies of other players \cite{PreDys2012,McAHau2016,UedTan2020}.
Therefore, a fair ZD strategy is an unbeatable strategy.
It should be noted that the left-hand side of Eq. (\ref{eq:fairZDS}) can be rewritten as $\sum_{a_{(1, j)}} c_{a_{(1, j)}} \left[ T_{(1, j)} \left( a_{(1, j)} | \bm{a}^\prime \right) - \delta_{a_{(1, j)}, a_{(1, j)}^\prime} \right]$.
Since the quantity $\delta_{a_{(1, j)}, a_{(1, j)}^\prime}$ can be regarded as the Repeat strategy, which repeats its own action, the left-hand side is a linear combination of the difference between a memory-one strategy $T_{(1, j)}$ and the Repeat strategy.
The fact that the time average of this quantity is zero is known as the Akin's lemma \cite{Aki2016}.

Below we note $B\left( \bm{a} \right):= s_j\left( \bm{a}_1 \right) - s_j\left( \bm{a}_2 \right)$.
A necessary and sufficient condition for the existence of a fair ZD strategy is obtained by applying the general existence condition of ZD strategies \cite{Ued2022b} to our problem.
\begin{proposition}
\label{prop:existence}
A fair ZD strategy of player $(1, j)$ against player $(2, j)$ exists if and only if there exist two different actions $\overline{a}_j, \underline{a}_j \in A_j$ such that
\begin{align}
 B \left( \left( \overline{a}_j, a_{(1, -j)} \right), \bm{a}_2 \right) &\geq 0 \quad \left( \forall a_{(1, -j)} \in A_{-j}, \forall \bm{a}_2 \in \mathcal{A} \right) \nonumber \\
 B \left( \left( \underline{a}_j, a_{(1, -j)} \right), \bm{a}_2 \right) &\leq 0 \quad \left( \forall a_{(1, -j)} \in A_{-j}, \forall \bm{a}_2 \in \mathcal{A} \right),
 \label{eq:condition_exsitence}
\end{align}
and $B$ is not identically zero.
\end{proposition}
It should be noted that, although the situation investigated in this paper is two concurrent repeated games, we can apply the result for repeated games to our problem if we regard $G=(G_1, G_2)$ as one repeated game.

\section{Results}
\label{sec:results}

\subsection{Existence condition of a fair ZD strategy}
Although the existence condition of a fair ZD strategy is generally given by Proposition \ref{prop:existence}, its implication is unclear.
In this subsection, we rewrite the condition in terms of the stage game $G$.
First, we introduce the concept of weakly payoff-monotonic games.
\begin{definition}
\label{def:weakPM}
A stage game $G$ is a \emph{weakly payoff-monotonic game} for player $(\mu, j)$ if there exist two actions $a_j^{(\mathrm{H})}, a_j^{(\mathrm{L})} \in A_j$ such that
\begin{align}
 s_j \left( a_j^{(\mathrm{H})}, a_{(\mu, -j)} \right) &= r^{(\mathrm{H})} \quad (\forall a_{(\mu, -j)} \in A_{-j}) \nonumber \\
 s_j \left( a_j^{(\mathrm{L})}, a_{(\mu, -j)} \right) &= r^{(\mathrm{L})} \quad (\forall a_{(\mu, -j)} \in A_{-j})
\end{align}
with some constants $r^{(\mathrm{H})}\geq r^{(\mathrm{L})}$ and they satisfy
\begin{align}
 r^{(\mathrm{L})} &\leq s_j \left( \bm{a}_\mu \right) \leq r^{(\mathrm{H})} \quad (\forall \bm{a}_\mu \in \mathcal{A}).
\end{align}
\end{definition}
Intuitively, a weakly payoff-monotonic game is a game where the strongest action and the weakest action exist and the payoffs of the two actions are independent of the opponents' actions.
If $s_j$ is a constant function, we call such a weakly payoff-monotonic game \emph{trivial} one.
Otherwise, we call a weakly payoff-monotonic game \emph{nontrivial} one.

For example, consider a two-player $4\times 2$ game with the payoff of player $(\mu, 1)$ given in Table \ref{table:four-two}.
\begin{table}[tb]
  \centering
  \caption{Payoff $s_1$ in a weakly payoff-monotonic game for player $(\mu, 1)$.}
  \begin{tabular}{|c|cc|} \hline
    & $d_1$ & $d_2$ \\ \hline
   $b_1$ & $1$ & $1$ \\
   $b_2$ & $1$ & $-1$ \\
   $b_3$ & $-1$ & $1$ \\
   $b_4$ & $-1$ & $-1$ \\ \hline
  \end{tabular}
  \label{table:four-two}
\end{table}
This game is a nontrivial weakly payoff-monotonic game for player $(\mu, 1)$ with $a_1^{(\mathrm{H})}=b_1$ and $a_1^{(\mathrm{L})}=b_4$.

We now prove our first main result.
\begin{theorem}
\label{thm:fairZD}
A fair ZD strategy of player $(1, j)$ against player $(2, j)$ exists if and only if the stage game is a nontrivial weakly payoff-monotonic game for player $(1, j)$.
\end{theorem}
See Section \ref{sec:proof_fairZD} for the proof.
Intuitively, $\overline{a}_j=a_j^{(\mathrm{H})}$ and $\underline{a}_j=a_j^{(\mathrm{L})}$ are the strongest action and the weakest action, respectively, and therefore player $(1, j)$ can unilaterally set $\mathcal{S}_{(1, j)}$ to any value in the interval $\left[ r^{(\mathrm{L})}, r^{(\mathrm{H})} \right]$.
A fair ZD strategy switches these two actions in order to enforce $\mathcal{S}_{(1, j)}=\mathcal{S}_{(2, j)}$ \cite{McAHau2016,Ued2022b}.

\begin{figure}[tbp]
\begin{center}
\includegraphics[clip, width=16.0cm]{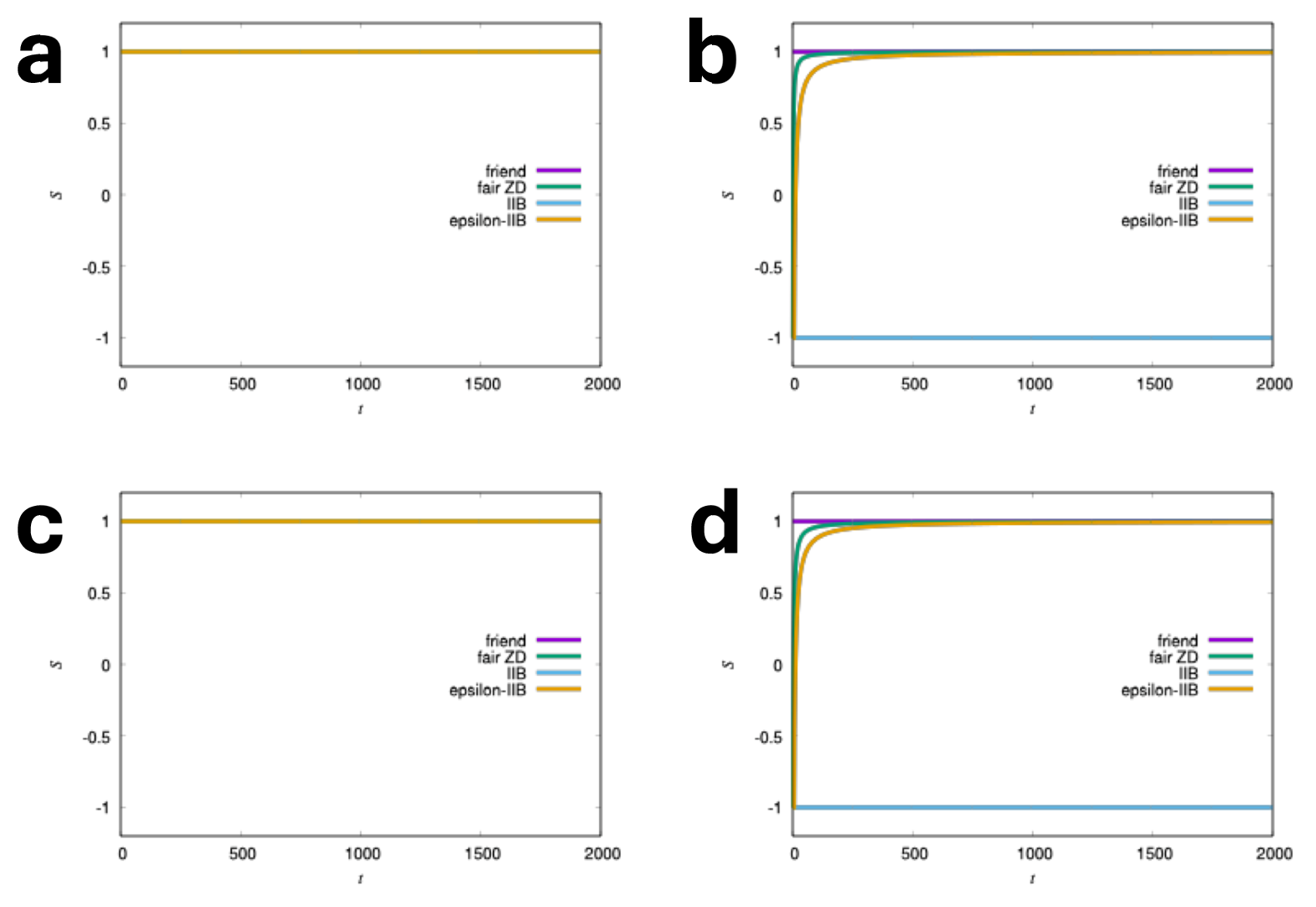}
\end{center}
\caption{The time-averaged payoffs of the focal player $(1, 1)$ and a friend $(2, 1)$ for the weakly payoff-monotonic game in Table \ref{table:four-two}. Players $(1,2)$, $(2,1)$, and $(2,2)$ take fixed actions $a_{(1,2)}=d_2$, $a_{(2,1)}=b_2$, and $a_{(2,2)}=d_1$. Player $(1,1)$ adopts three strategies: a fair ZD strategy \cite{Ued2022b}, IIB (\ref{eq:IIB}), and $\epsilon$-IIB (\ref{eq:eIIB}) with $\epsilon=0.01$. The initial conditions are (a) $a_{(1,1)}=b_1$, (b) $a_{(1,1)}=b_2$, (c) $a_{(1,1)}=b_3$, and (d) $a_{(1,1)}=b_4$. The time-averaged payoffs are calculated by one sample. In panels (a) and (c), all curves overlap.}
\label{fig:weakly}
\end{figure}

In order to check the validity of this result, we have performed a numerical simulation for the game in Table \ref{table:four-two}.
Players $(1,2)$, $(2,1)$, and $(2,2)$ take fixed actions $a_{(1,2)}=d_2$, $a_{(2,1)}=b_2$, and $a_{(2,2)}=d_1$, and player $(1,1)$ adopts a fair ZD strategy.
A fair ZD strategy is constructed by the method in \cite{Ued2022b}.
Four initial conditions of $a_{(1,1)}$ are used in the simulation.
The payoffs are calculated by the time-averaged payoffs of one sample.
The result is displayed in Fig. \ref{fig:weakly}.
We find that the payoff of a friend $(2,1)$ is $\mathcal{S}_{(2, 1)}=1$ and the payoff $\mathcal{S}_{(1, 1)}$ seems to converge to $1$.

\subsection{Existence condition of unbeatable imitation strategies}
In this subsection, we investigate the condition in which imitation is unbeatable.
Here, we introduce the concept of strongly payoff-monotonic games.
\begin{definition}
\label{def:strongPM}
A stage game $G$ is a \emph{strongly payoff-monotonic game} for player $(\mu, j)$ if the payoff $s_j$ satisfies
\begin{align}
 s_j\left( a_{(\mu, j)}, a_{(\mu, -j)} \right) &= r\left( a_{(\mu, j)} \right) \quad \left( \forall a_{(\mu, j)} \in A_j, \forall a_{(\mu, -j)} \in A_{-j} \right)
\end{align}
with some function $r$.
\end{definition}
Intuitively, a strongly payoff-monotonic game is a game where the order of strength of actions can be defined and the payoff of each action is independent of the opponents' actions.
Similarly as before, if $s_j$ is a constant function, we call such a strongly payoff-monotonic game \emph{trivial} one.
Otherwise, we call a strongly payoff-monotonic game \emph{nontrivial} one.
By definition, a strongly payoff-monotonic game for player $(\mu, j)$ is a weakly payoff-monotonic game for player $(\mu, j)$.

We prove our second main result.
\begin{theorem}
\label{thm:IIB}
IIB of player $(1, j)$ is unbeatable against player $(2, j)$ if and only if the stage game is a strongly payoff-monotonic game for player $(1, j)$.
\end{theorem}
See Section \ref{sec:proof_IIB} for the proof.
Intuitively, if the actions $a_j \in A_j$, $a_{-j}, a_{-j}^\prime \in A_{-j}$ such that $s_j\left( a_j, a_{-j} \right) < s_j\left( a_j, a_{-j}^\prime \right)$ exist, infinite repetition of the action profiles $\bm{a}_1 = (a_j, a_{-j})$ and $\bm{a}_2 = (a_j, a_{-j}^\prime)$ can be realized by IIB of player $(1, j)$, who has the freedom of choosing any initial action, which leads to $\mathcal{S}_{(1, j)} < \mathcal{S}_{(2, j)}$.

Furthermore, a similar result is obtained for TFT.
\begin{theorem}
\label{thm:TFT}
TFT of player $(1, j)$ is unbeatable against player $(2, j)$ if and only if the stage game is a strongly payoff-monotonic game for player $(1, j)$.
\end{theorem}
See Section \ref{sec:proof_TFT} for the proof.
The proof is almost the same as that for Theorem \ref{thm:IIB}.

It is noteworthy that IIB and TFT of player $(1, 1)$ is beaten by player $(2, 1)$ in the weakly payoff-monotonic game given in Table \ref{table:four-two}, when $\bm{a}_1=(b_2, d_2)$ and $\bm{a}_2=(b_2, d_1)$ are infinitely repeated.
In fact, in Fig. \ref{fig:weakly}b, we can see that the payoff of IIB is $\mathcal{S}_{(1, 1)}=-1$ whereas the payoff of a friend $(2,1)$ is $\mathcal{S}_{(2, 1)}=1$.
In contrast, a two-player $3\times 2$ game with the payoff of player $(\mu, 1)$ given in Table \ref{table:three-two} is a strongly payoff-monotonic game for player $(\mu, 1)$.
\begin{table}[tb]
  \centering
  \caption{Payoff $s_1$ in a strongly payoff-monotonic game for player $(\mu, 1)$.}
  \begin{tabular}{|c|cc|} \hline
    & $d_1$ & $d_2$ \\ \hline
   $b_1$ & $1$ & $1$ \\
   $b_2$ & $0$ & $0$ \\
   $b_3$ & $-1$ & $-1$ \\ \hline
  \end{tabular}
  \label{table:three-two}
\end{table}
For this game, both IIB and TFT are unbeatable.

\begin{figure}[tbp]
\begin{center}
\includegraphics[clip, width=16.0cm]{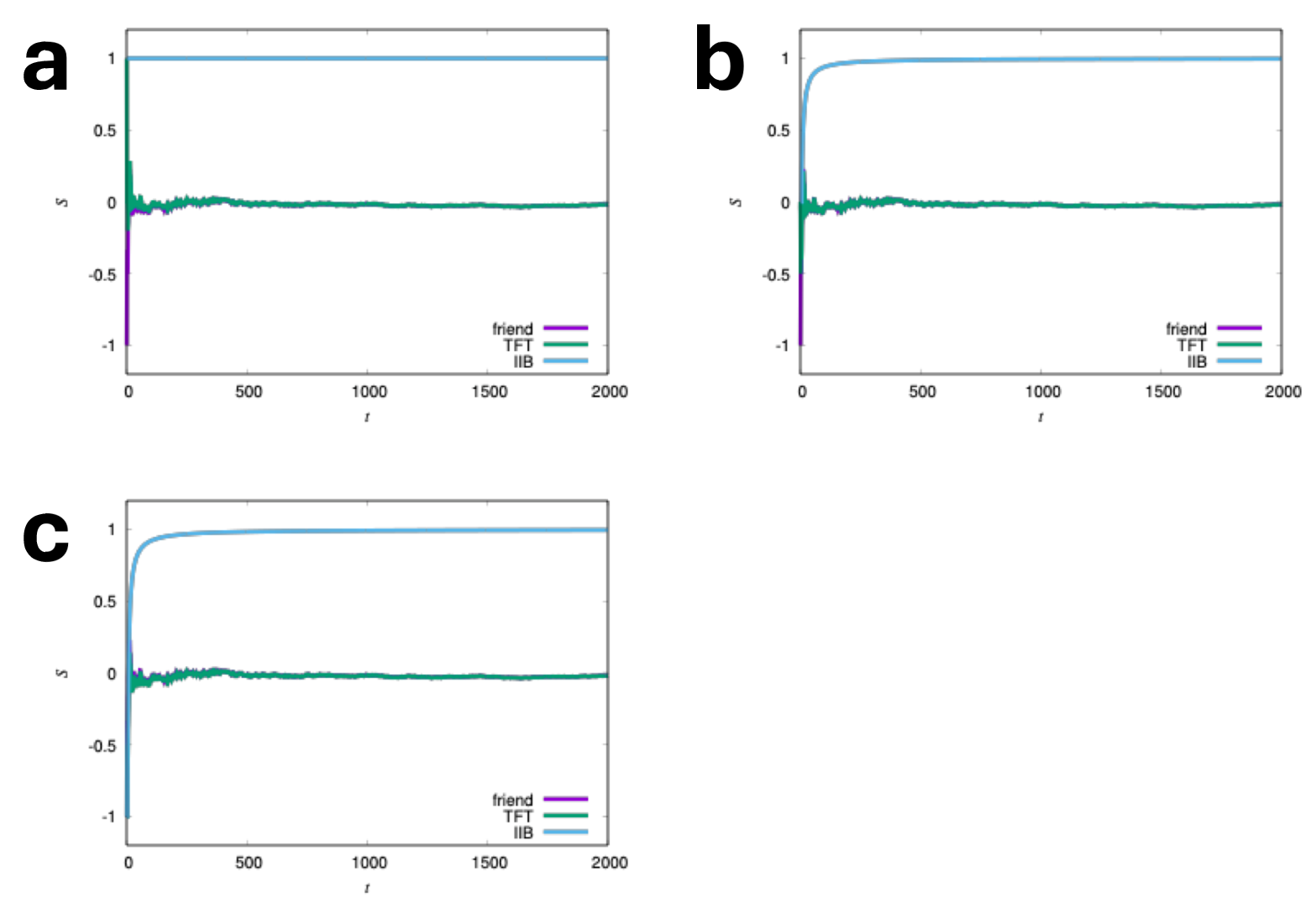}
\end{center}
\caption{The time-averaged payoffs of the focal player $(1, 1)$ and a friend $(2, 1)$ for the strongly payoff-monotonic game in Table \ref{table:three-two}. Players $(1,2)$, $(2,1)$, and $(2,2)$ take random actions in each round. Player $(1,1)$ adopts two strategies: TFT (\ref{eq:TFT}) and IIB (\ref{eq:IIB}). The initial condition of the focal player is (a) $a_{(1,1)}=b_1$, (b) $a_{(1,1)}=b_2$, and (c) $a_{(1,1)}=b_3$. The initial condition of the other players is $a_{(1,2)}=d_1$, $a_{(2,1)}=b_3$, and $a_{(2,2)}=d_1$. The time-averaged payoffs are calculated by one sample.}
\label{fig:strongly}
\end{figure}

In order to check the validity of these results, we have performed a numerical simulation for the game in Table \ref{table:three-two}.
Players $(1,2)$, $(2,1)$, and $(2,2)$ take random actions in each round, and player $(1,1)$ adopts TFT or IIB.
The initial condition of players $(1,2)$, $(2,1)$, and $(2,2)$ is $a_{(1,2)}=d_1$, $a_{(2,1)}=b_3$, and $a_{(2,2)}=d_1$.
Three initial conditions of $a_{(1,1)}$ are used in the simulation.
The payoffs are calculated by the time-averaged payoffs of one sample.
The result is displayed in Fig. \ref{fig:strongly}.
We find that both TFT and IIB of player $(1,1)$ are unbeatable against a friend $(2,1)$.
Particularly, whereas TFT seems to satisfy $\mathcal{S}_{(1, 1)}=\mathcal{S}_{(2, 1)}$, IIB satisfies $\mathcal{S}_{(1, 1)}\geq \mathcal{S}_{(2, 1)}$, which is consistent with the proofs.

It should be noted that stronger versions of Theorems \ref{thm:IIB} and \ref{thm:TFT}, which do not assume the existence of Eq. (\ref{eq:payoff_repeated}), can also be obtained as in \ref{app:stronger_IIBTFT}.

\subsection{Unbeatable imitation as ZD strategies}
The proof of Theorem \ref{thm:TFT} implies that TFT of player $(1, j)$ unilaterally enforces $\mathcal{S}_{(1, j)} = \mathcal{S}_{(2, j)}$ in strongly payoff-monotonic games for player $(1, j)$, and suggests that TFT of player $(1, j)$ may be a fair ZD strategy.
Here we prove that this expectation is true.
\begin{theorem}
\label{thm:TFT_ZDS}
If the stage game is a nontrivial strongly payoff-monotonic game for player $(1, j)$, TFT of player $(1, j)$ is a fair ZD strategy against player $(2, j)$.
\end{theorem}
See Section \ref{sec:proof_TFT_ZDS} for the proof.
Intuitively, this result comes from the fact that the quantity $s_j\left( \bm{a}_1 \right) - s_j\left( \bm{a}_2 \right)$ depends only on $a_{(1, j)}$ and $a_{(2, j)}$ in a strongly payoff-monotonic game.

A slightly different result is obtained for IIB.
For this purpose, we write the probability distribution of action profiles at the $t$-th round by $P_t(\bm{a})$, and introduce the time-averaged distribution \cite{Aki2016,Ued2022}
\begin{align}
 P^*(\bm{a}) &:= \lim_{T\rightarrow \infty} \frac{1}{T} \sum_{t=1}^T P_t(\bm{a}).
\end{align}
We write an expected value with respect to $P^*$ by $\left\langle \cdots \right\rangle^*$.
By using this notation, $\mathcal{S}_{(\mu, j)}=\left\langle s_{(\mu, j)} \right\rangle^*$, for example.
Generally, if a memory-one strategy $T_{(\mu, j)}$ of player $(\mu, j)$ is described in the form
\begin{align}
 \sum_{a_{(\mu, j)}} c_{a_{(\mu, j)}} T_{(\mu, j)} \left( a_{(\mu, j)} | \bm{a}^\prime \right) - c_{a_{(\mu, j)}^\prime} &= D\left( \bm{a}^\prime \right) \quad \left( \forall \bm{a}^\prime \in \mathcal{A} \times \mathcal{A} \right)
 \label{eq:ZDS}
\end{align}
with some function $D$ and some coefficients $\left\{ c_{a_{(\mu, j)}}  \right\}$ and both sides are not identically zero, it is called a ZD strategy unilaterally enforcing $\left\langle D \right\rangle^*=0$ \cite{Ued2025}.
\begin{theorem}
\label{thm:IIB_ZDS}
If the stage game is a nontrivial strongly payoff-monotonic game for player $(1, j)$, IIB of player $(1, j)$ is a ZD strategy which unilaterally enforces
\begin{align}
 \left\langle \left[ s_j\left( \bm{a}_2^\prime \right) - s_j\left( \bm{a}_1^\prime \right) \right] \mathbb{I}\left( s_j\left( \bm{a}_1^\prime \right) < s_j\left( \bm{a}_2^\prime \right) \right) \right\rangle^* &= 0.
 \label{eq:linear_IIB}
\end{align}
\end{theorem}
See Section \ref{sec:proof_IIB_ZDS} for the proof.
The proof is almost the same as that for Theorem \ref{thm:TFT_ZDS}.
We remark that Eq. (\ref{eq:linear_IIB}) is rewritten as
\begin{align}
 \sum_{\bm{a}}^{s_j\left( \bm{a}_1 \right) < s_j\left( \bm{a}_2 \right)} \left[ s_j\left( \bm{a}_2 \right) - s_j\left( \bm{a}_1 \right) \right] P^*(\bm{a}) &= 0,
\end{align}
and implies that $P^*(\bm{a})=0$ for all $\bm{a}$ such that $s_j\left( \bm{a}_1 \right) < s_j\left( \bm{a}_2 \right)$ \cite{Ued2022,Ued2022c}.
Therefore, Eq. (\ref{eq:linear_IIB}) means that IIB is an unbeatable ZD strategy, which is consistent with Theorem \ref{thm:IIB}.

It should be noted that, if a strongly payoff-monotonic game for player $(1, j)$ is a trivial one, $s_j$ is a constant function, and both TFT and IIB of player $(1, j)$ are trivially unbeatable against player $(2, j)$.
However, for such a case, they are not regarded as ZD strategies, since they do not satisfy the definition (\ref{eq:ZDS}) of ZD strategies.

\subsection{Improvement of Imitate-If-Better strategy}
Theorem \ref{thm:IIB} states that IIB is unbeatable if and only if the stage game is a strongly payoff-monotonic game.
In this subsection, we consider an improvement of IIB, which is unbeatable even in weakly payoff-monotonic games.
\begin{definition}
\label{def:eIIB}
A memory-one strategy $T_{(1, j)}$ of player $(1, j)$ is the \emph{$\epsilon$-Imitate-If-Better ($\epsilon$-IIB)} strategy against player $(2, j)$ if it is described as
\begin{align}
 T_{(1, j)} \left( a_{(1, j)} | \bm{a}^\prime \right) &= \left[ (1-\epsilon) \delta_{a_{(1, j)}, a_{(2, j)}^\prime} + \epsilon \frac{1}{|A_j|} \right] \mathbb{I} \left( s_j\left( \bm{a}_1^\prime \right) < s_j\left( \bm{a}_2^\prime \right) \right) \nonumber \\
 & \quad + \delta_{a_{(1, j)}, a_{(1, j)}^\prime} \mathbb{I} \left( s_j\left( \bm{a}_1^\prime \right) \geq s_j\left( \bm{a}_2^\prime \right) \right)
 \label{eq:eIIB}
\end{align}
with $0< \epsilon \leq 1$.
\end{definition}
That is, $\epsilon$-IIB randomly samples an action with probability $\epsilon$ when it was beaten in the previous round.
This strategy including exploration seems to be similar to $\epsilon$-greedy method in reinforcement learning \cite{SutBar2018}, although $\epsilon$-IIB does not contain learning.
In order to explain the property of $\epsilon$-IIB, we introduce the following class of games.
\begin{definition}
\label{def:GAA}
A stage game $G$ is a \emph{game with an autocratic action} for player $(\mu, j)$ if there exists an action $a_j^{(\mathrm{H})} \in A_j$ such that
\begin{align}
 s_j \left( a_j^{(\mathrm{H})}, a_{(\mu, -j)} \right) &= r^{(\mathrm{H})} \quad (\forall a_{(\mu, -j)} \in A_{-j})
\end{align}
with some constant $r^{(\mathrm{H})}$ and it satisfies
\begin{align}
 s_j \left( \bm{a}_\mu \right) &\leq r^{(\mathrm{H})} \quad (\forall \bm{a}_\mu \in \mathcal{A}).
\end{align}
\end{definition}
For example, the game in Table \ref{table:GAA} is a game with an autocratic action.
\begin{table}[tb]
  \centering
  \caption{Payoff $s_1$ in a game with an autocratic action.}
  \begin{tabular}{|c|cc|} \hline
    & $d_1$ & $d_2$ \\ \hline
   $b_1$ & $1$ & $1$ \\
   $b_2$ & $1$ & $-1$ \\
   $b_3$ & $-1$ & $1$ \\ \hline
  \end{tabular}
  \label{table:GAA}
\end{table}
Trivially, a weakly payoff-monotonic game is a game with an autocratic action.
It should be noted that, although the term ``autocratic action'' is inspired by ``autocratic strategy'' \cite{McAHau2016}, they are different concepts and one should not confuse the two notions.
An autocratic action is precisely a dominant action independent of the opponents' actions.

We can prove the following theorem for $\epsilon$-IIB.
\begin{theorem}
\label{thm:eIIB}
$\epsilon$-IIB of player $(1, j)$ is unbeatable almost surely against player $(2, j)$ if the stage game is a game with an autocratic action for player $(1, j)$.
\end{theorem}
See Section \ref{sec:proof_eIIB} for the proof.
Intuitively, every time $\epsilon$-IIB is beaten, it can choose the strongest action $a_j^{(\mathrm{H})}$ with positive probability.
Therefore, when we just add exploration to IIB, it becomes unbeatable even in games with an autocratic action, including weakly payoff-monotonic games.
In fact, we can see in Fig. \ref{fig:weakly} that, although IIB of player $(1,1)$ can be beaten by a friend $(2,1)$, $\epsilon$-IIB of player $(1,1)$ performs as well as a friend $(2,1)$.

\begin{figure}[tbp]
\begin{center}
\includegraphics[clip, width=16.0cm]{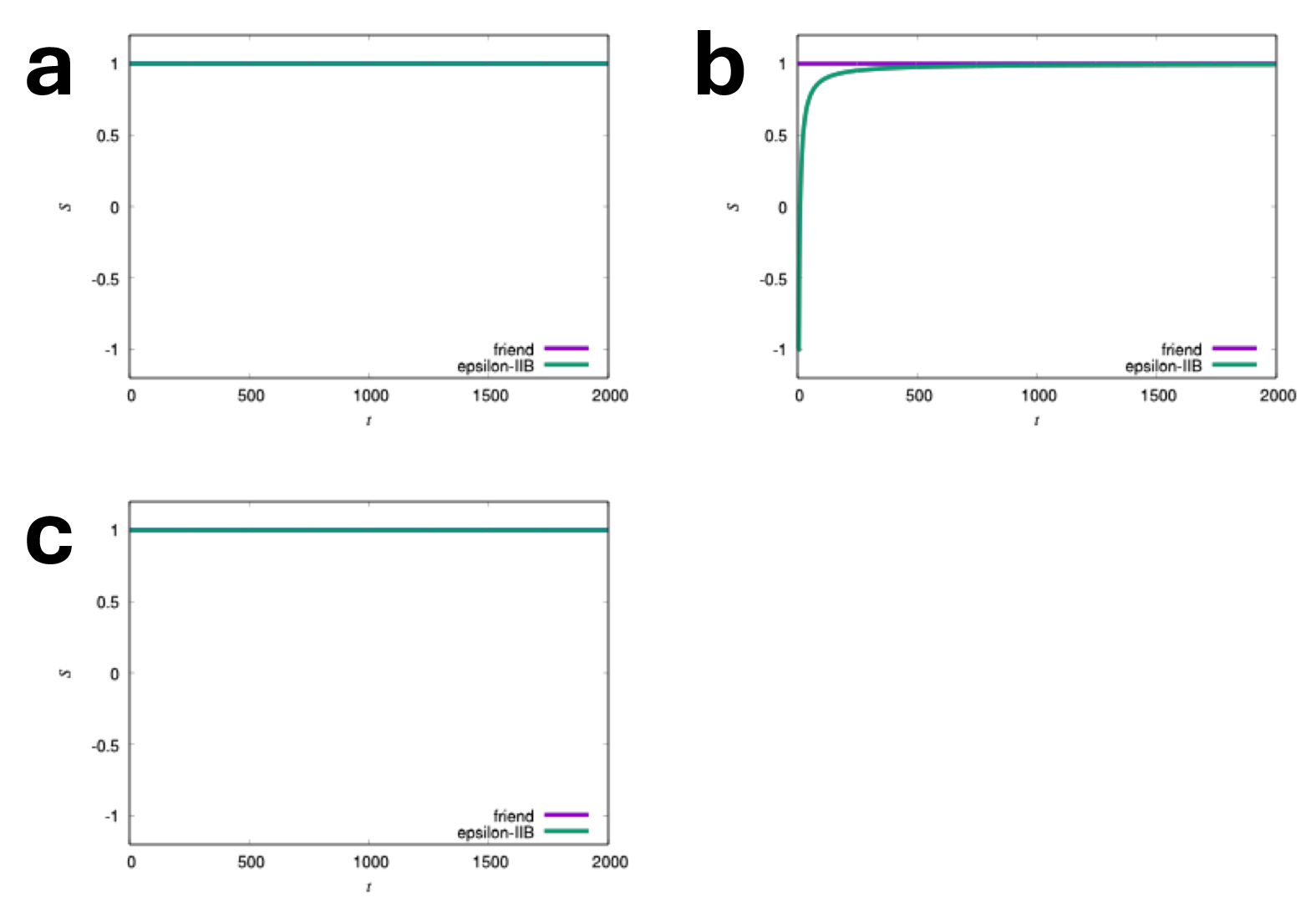}
\end{center}
\caption{The time-averaged payoffs of the focal player $(1, 1)$ and a friend $(2, 1)$ for the game with an autocratic action in Table \ref{table:GAA}. Players $(1,2)$, $(2,1)$, and $(2,2)$ take fixed actions $a_{(1,2)}=d_2$, $a_{(2,1)}=b_2$, and $a_{(2,2)}=d_1$. Player $(1,1)$ adopts $\epsilon$-IIB (\ref{eq:eIIB}) with $\epsilon=0.01$. The initial conditions are (a) $a_{(1,1)}=b_1$, (b) $a_{(1,1)}=b_2$, and (c) $a_{(1,1)}=b_3$. The time-averaged payoffs are calculated by one sample. In panels (a) and (c), all curves overlap.}
\label{fig:GAA}
\end{figure}

Furthermore, we have performed a numerical simulation for the game in Table \ref{table:GAA}.
Players $(1,2)$, $(2,1)$, and $(2,2)$ take fixed actions $a_{(1,2)}=d_2$, $a_{(2,1)}=b_2$, and $a_{(2,2)}=d_1$, and player $(1,1)$ adopts $\epsilon$-IIB (\ref{eq:eIIB}) with $\epsilon=0.01$.
Three initial conditions of $a_{(1,1)}$ are used in the simulation.
The payoffs are calculated by the time-averaged payoffs of one sample.
The result is consistent with Theorem \ref{thm:eIIB} and is displayed in Fig. \ref{fig:GAA}.

\section{Discussion}
\label{sec:discussion}
In this paper, we specified the existence conditions for unbeatable imitation (Theorems \ref{thm:IIB} and \ref{thm:TFT}) and a fair ZD strategy (Theorem \ref{thm:fairZD}) in the ``imitation of friends'' situations.
As expected, a fair ZD strategy exists under a weaker condition than unbeatable imitation strategies, similarly to the ``imitation of opponents'' situations \cite{Ued2022,Ued2022b,Ued2023}.
In the ``imitation of opponents'' situations, unbeatable imitation exists when the order of strength of actions exists, and a fair ZD strategy exists when the strongest action and the weakest action exist \cite{Ued2023}.
This property also holds in our ``imitation of friends'' situations.
However, in the ``imitation of friends'' situations, these existence conditions strongly restrict the structure of stage games, compared with the ``imitation of opponents'' situations.
In fact, weakly payoff-monotonic games and strongly payoff-monotonic games describe situations in which a player can unilaterally obtain the largest payoff regardless of actions of other players, which are almost nonsense situations.
Therefore, success of imitation strategies in the ``imitation of friends'' situations seems to be very limited.

Although weakly payoff-monotonic games and strongly payoff-monotonic games may be almost nonsense situations, rather our results are useful when a player does not have complete knowledge about payoff functions.
For example, let us regard the game in Table \ref{table:three-two} as a three-armed bandit \cite{Sch1998}.
Player $(\mu, 1)$ is one playing the bandit, and player $(\mu, 2)$ is the state of the bandit (but has no influence on the outcomes).
Whereas the best action for player $(\mu, 1)$ is $b_1$, we assume that player $(1, 1)$ initially does not completely know the payoff function $s_1$ but knows the fact that player $(2, 1)$ plays the same bandit as player $(1, 1)$, and observes the previous action and the obtained payoff of player $(2, 1)$ every round.
Then, our results ensure that the imitation strategies of player $(1, 1)$ obtain the time-averaged payoff not less than that of player $(2, 1)$, even though player $(1, 1)$ does not completely know the payoff function.
Therefore, our results characterize success of imitation regardless of incomplete knowledge about payoff functions.
An additional numerical result related to this point is provided in \ref{app:bandit}.

Of course, the limitation on success of imitation comes from the fact that we only consider the situations where an imitator naively imitates the previous action of a friend, regardless of the previous actions of the opponents.
If we consider more clever imitation which imitates others only on some specific conditions, situations may become better.
In fact, Theorem \ref{thm:eIIB} claims that IIB with exploration can be unbeatable even in weakly payoff-monotonic games.
Exploration and exploitation are also on the basis of reinforcement learning \cite{SutBar2018}.
In reinforcement learning, a player learns its optimal strategy based on its own experience, not by imitating actions of other players.
Concretely, the $\epsilon$-greedy method in reinforcement learning is described as
\begin{align}
 T_{(1, j)} \left( a_{(1, j)} | \bm{a}^\prime \right) &= (1-\epsilon) \delta_{a_{(1, j)}, \arg\max_a Q_{(1,j)} \left( a, \bm{a}^\prime \right)} + \epsilon \frac{1}{|A_j|},
\end{align}
where $Q_{(1,j)}\left( a_{(1, j)}, \bm{a}^\prime \right):=\mathbb{E} \left[ \left. \sum_{t^\prime=0}^\infty \delta^{t^\prime} s_j\left( \bm{a}_1^{(t+t^\prime+1)} \right) \right| a_{(1, j)}^{(t+1)} = a_{(1, j)}, \bm{a}^{(t)}=\bm{a}^\prime \right]$ with a discount factor $\delta\in [0,1)$ is called the action-value function and characterizes the expected future payoff obtained by action $a_{(1, j)}$ when the previous action profile was $\bm{a}^\prime$.
In contrast, our $\epsilon$-IIB usually imitates an action of a friend.
It has been known that individual learning is also important as well as social learning when environmental changes occur \cite{WAF2004}.
Although imitation is one form of social learning, it is too myopic, and incorporating the effect of individual learning may make imitation more unbeatable.

We also found that the unbeatable imitation strategies can be interpreted as unbeatable ZD strategies in the ``imitation of friends'' situations (Theorems \ref{thm:TFT_ZDS} and \ref{thm:IIB_ZDS}).
Similar results were obtained for the ``imitation of opponents'' situations \cite{Ued2022}.
In the two-player ``imitation of opponents'' situations, TFT and IIB are unbeatable ZD strategies if the stage game is a nontrivial potential game \cite{MonSha1996}, where the order of strength of actions can be defined.
Therefore, Theorems \ref{thm:TFT_ZDS} and \ref{thm:IIB_ZDS} can be regarded as an extension of the previous results to ``imitation of friends'' situations, and confirm our conjecture that unbeatable property of imitation is realized by a payoff-control ability of ZD strategies.

From a perspective of ZD strategies, fair ZD strategies are just one class of unbeatable ZD strategies.
In unbeatable ZD strategies, there also exist the extortionate ZD strategies, which unilaterally obtain payoffs not less than those of opponents \cite{PreDys2012,HNS2013,AdaHin2013,HWTN2014,McAHau2016}.
Therefore, unbeatable ZD strategies may exist under weaker conditions than fair ZD strategies, although fair ZD strategies are easier to understand \cite{MMHet2025}.
Moreover, unbeatable strategies are not limited to ZD strategies, and there exist more general memory-one strategies which unilaterally control payoffs into some regions in social dilemma games \cite{HTS2015,Aki2016}.
Specifying a necessary and sufficient condition for the existence of unbeatable strategies in repeated games is a significant open problem.

In addition, it should be noted that we have implicitly used several assumptions.
For example, we assumed that the game is perfect-monitoring, and there are no observation errors.
In more realistic situations, previous action profiles may be imperfectly observed \cite{HRZ2015,UedTan2020,MamIch2020}.
For the repeated prisoner's dilemma game, it was known that, although TFT is a fair ZD strategy in the perfect-monitoring case, it is generally not a fair ZD strategy when observation errors exist \cite{MamIch2019}.
We conjecture that unbeatable imitation strategies in the ``imitation of a friend'' situations in this paper may not be unbeatable if observation errors exist.
Moreover, how other factors such as structured populations \cite{SAP2022}, individual sensitivity \cite{WFTW2023}, and environmental fluctuations \cite{WCTZ2024}, have effects on our results should be clarified in future.

Last but not least, we remark that, in this paper, we have not assumed any symmetries in each $G_\mu$, although we assumed that $G_1$ and $G_2$ have the same structure.
If we assume some symmetries in each $G_\mu$, players can implement imitation strategies using information of both opponents and friends \cite{AHO2007}.
For such situations, whereas we can use previous results for the ``imitation of opponents'' situations \cite{DOS2012b,DOS2014,Ued2022,Ued2022b,Ued2023}, the existence conditions of unbeatable imitation and unbeatable ZD strategies will be more complicated.
Together with the ``imitation of friends'' situations where the number of parallel games is more than two, such more general situations should be investigated in future.

\section{Methods}
\label{sec:methods}

\subsection{Proof of Theorem \ref{thm:fairZD}}
\label{sec:proof_fairZD}
According to Proposition \ref{prop:existence}, a fair ZD strategy of player $(1, j)$ exists if and only if there exist two different actions $\overline{a}_j, \underline{a}_j \in A_j$ such that
\begin{align}
 s_j\left( \overline{a}_j, a_{(1, -j)} \right) - s_j\left( \bm{a}_2 \right) &\geq 0 \quad \left( \forall a_{(1, -j)} \in A_{-j}, \forall \bm{a}_2 \in \mathcal{A} \right) \nonumber \\
 s_j\left( \underline{a}_j, a_{(1, -j)} \right) - s_j\left( \bm{a}_2 \right) &\leq 0 \quad \left( \forall a_{(1, -j)} \in A_{-j}, \forall \bm{a}_2 \in \mathcal{A} \right),
 \label{eq:existence_fair}
\end{align}
and $s_j$ is not a constant function.

(Necessity.)
Suppose that a fair ZD strategy of player $(1, j)$ exists.
If Eq. (\ref{eq:existence_fair}) holds for some $\overline{a}_j, \underline{a}_j \in A_j$, we particularly obtain
\begin{align}
 s_j\left( \overline{a}_j, a_{(1, -j)} \right) - s_j\left( \overline{a}_j, a_{(2, -j)} \right) &\geq 0 \quad \left( \forall a_{(1, -j)}, \forall a_{(2, -j)} \in A_{-j} \right) \nonumber \\
 s_j\left( \underline{a}_j, a_{(1, -j)} \right) - s_j\left( \underline{a}_j, a_{(2, -j)} \right) &\leq 0 \quad \left( \forall a_{(1, -j)}, \forall a_{(2, -j)} \in A_{-j} \right)
\end{align}
for $a_{(2,j)}=\overline{a}_j$ and $a_{(2,j)}=\underline{a}_j$, respectively.
However, because both $a_{(1, -j)}$ and $a_{(2, -j)}$ are contained in the same set $A_{-j}$, we find
\begin{align}
 s_j\left( \overline{a}_j, a_{(1, -j)} \right) &= \mathrm{constant} =: \overline{r} \quad \left( \forall a_{(1, -j)} \in A_{-j} \right) \nonumber \\
 s_j\left( \underline{a}_j, a_{(1, -j)} \right) &= \mathrm{constant} =: \underline{r} \quad \left( \forall a_{(1, -j)} \in A_{-j} \right).
\end{align}
By using this result in Eq. (\ref{eq:existence_fair}), we obtain
\begin{align}
 \overline{r} - s_j\left( \bm{a}_2 \right) &\geq 0 \quad \left( \forall \bm{a}_2 \in \mathcal{A} \right) \nonumber \\
 \underline{r} - s_j\left( \bm{a}_2 \right) &\leq 0 \quad \left( \forall \bm{a}_2 \in \mathcal{A} \right),
\end{align}
which implies that the stage game is a weakly payoff-monotonic game for player $(1, j)$ with $a_j^{(\mathrm{H})} = \overline{a}_j$, $a_j^{(\mathrm{L})}=\underline{a}_j$, $r^{(\mathrm{H})}=\overline{r}$, and $r^{(\mathrm{L})}=\underline{r}$.
Furthermore, because $s_j$ is not a constant function, this weakly payoff-monotonic game is nontrivial one.

(Sufficiency.)
If the stage game is a nontrivial weakly payoff-monotonic game for player $(1, j)$, we obtain 
\begin{align}
 s_j\left( a_j^{(\mathrm{H})}, a_{(1, -j)} \right) - s_j\left( \bm{a}_2 \right) &\geq 0 \quad \left( \forall a_{(1, -j)} \in A_{-j}, \forall \bm{a}_2 \in \mathcal{A} \right) \nonumber \\
 s_j\left( a_j^{(\mathrm{L})}, a_{(1, -j)} \right) - s_j\left( \bm{a}_2 \right) &\leq 0 \quad \left( \forall a_{(1, -j)} \in A_{-j}, \forall \bm{a}_2 \in \mathcal{A} \right),
\end{align}
which is equivalent to Eq. (\ref{eq:existence_fair}).
Furthermore, by assumption, $s_j$ is not a constant function.
Therefore, a fair ZD strategy of player $(1, j)$ exists.
$\Box$

\subsection{Proof of Theorem \ref{thm:IIB}}
\label{sec:proof_IIB}
(Necessity.)
Suppose that IIB of player $(1, j)$ is unbeatable against player $(2, j)$.
Assume that there exist $a_j \in A_j$, $a_{-j}, a_{-j}^\prime \in A_{-j}$ such that 
\begin{align}
 s_j\left( a_j, a_{-j} \right) - s_j\left( a_j, a_{-j}^\prime \right) &< 0.
\end{align}
When $\bm{a}_1 = (a_j, a_{-j})$ and $\bm{a}_2 = (a_j, a_{-j}^\prime)$, IIB of player $(1, j)$ continues to take $a_j$ in the next round.
However, if such an action profile is infinitely repeated, IIB of player $(1, j)$ is beaten by player $(2, j)$: $\mathcal{S}_{(1, j)} < \mathcal{S}_{(2, j)}$.
Therefore, for all $a_j\in A_j$, the equation
\begin{align}
 s_j\left( a_j, a_{-j} \right) &= r\left( a_j \right) \quad \left( \forall a_{-j} \in A_{-j} \right)
\end{align}
must hold with some function $r$.
This implies that the stage game is a strongly payoff-monotonic game for player $(1, j)$.

(Sufficiency.)
Suppose that the stage game is a strongly payoff-monotonic game for player $(1, j)$.
We can sort $r\left( a_{(1, j)} \right)$ in ascending order.
Because IIB of player $(1, j)$ takes $a_{(1, j)}^{(t)}$ at the $t$-th round with $t\geq 2$ such that
\begin{align}
 r\left( a_{(1, j)}^{(t)} \right) &= \max \left\{ r\left( a_{(1, j)}^{(1)} \right),  \max \left\{ r\left( a_{(2, j)}^{(t^\prime)} \right) | 1\leq t^\prime \leq t-1 \right\} \right\},
\end{align}
and because $A_j$ is a finite set, we obtain
\begin{align}
 \mathcal{S}_{(1, j)} &= \lim_{T\rightarrow \infty} \frac{1}{T} \sum_{t=1}^T \mathbb{E}\left[ s_j\left( \bm{a}_1^{(t)} \right) \right] \nonumber \\
 &= \lim_{T\rightarrow \infty} \frac{1}{T} \sum_{t=1}^T \mathbb{E}\left[  r\left( a_{(1, j)}^{(t)} \right) \right] \nonumber \\
 &= \lim_{T\rightarrow \infty} \frac{1}{T} \sum_{t=1}^T \mathbb{E}\left[  \max \left\{ r\left( a_{(1, j)}^{(1)} \right), \max \left\{ r\left( a_{(2, j)}^{(t^\prime)} \right) | 1\leq t^\prime \leq t-1 \right\} \right\} \right] \nonumber \\
 &\geq \lim_{T\rightarrow \infty} \frac{1}{T} \left[ \sum_{t=2}^T \mathbb{E}\left[ r\left( a_{(2, j)}^{(t-1)} \right) \right] + \mathbb{E}\left[ r\left( a_{(1, j)}^{(1)} \right) \right] \right] \nonumber \\
 &= \lim_{T\rightarrow \infty} \frac{1}{T} \left[ \sum_{t=1}^T \mathbb{E}\left[ r\left( a_{(2, j)}^{(t)} \right) \right] - \mathbb{E}\left[ r\left( a_{(2, j)}^{(T)} \right) \right] + \mathbb{E}\left[ r\left( a_{(1, j)}^{(1)} \right) \right] \right] \nonumber \\
 &= \mathcal{S}_{(2, j)}.
\end{align}
Therefore, IIB of player $(1, j)$ is unbeatable against player $(2, j)$.
$\Box$

\subsection{Proof of Theorem \ref{thm:TFT}}
\label{sec:proof_TFT}
(Necessity.)
Necessity is proved by the same method as that in the proof of Theorem \ref{thm:IIB}.

(Sufficiency.)
Suppose that the stage game is a strongly payoff-monotonic game for player $(1, j)$.
We can sort $r\left( a_{(1, j)} \right)$ in ascending order.
Because TFT of player $(1, j)$ takes
\begin{align}
 a_{(1, j)}^{(t)} &= a_{(2, j)}^{(t-1)}
\end{align}
at the $t$-th round with $t\geq 2$, and because $A_j$ is a finite set, we obtain
\begin{align}
 \mathcal{S}_{(1, j)} &= \lim_{T\rightarrow \infty} \frac{1}{T} \sum_{t=1}^T \mathbb{E}\left[ s_j\left( \bm{a}_1^{(t)} \right) \right] \nonumber \\
 &= \lim_{T\rightarrow \infty} \frac{1}{T} \sum_{t=1}^T \mathbb{E}\left[  r\left( a_{(1, j)}^{(t)} \right) \right] \nonumber \\
 &= \lim_{T\rightarrow \infty} \frac{1}{T} \left[ \sum_{t=2}^T \mathbb{E}\left[ r\left( a_{(2, j)}^{(t-1)} \right) \right] + \mathbb{E}\left[ r\left( a_{(1, j)}^{(1)} \right) \right] \right] \nonumber \\
 &= \lim_{T\rightarrow \infty} \frac{1}{T} \left[ \sum_{t=1}^T \mathbb{E}\left[ r\left( a_{(2, j)}^{(t)} \right) \right] - \mathbb{E}\left[ r\left( a_{(2, j)}^{(T)} \right) \right] + \mathbb{E}\left[ r\left( a_{(1, j)}^{(1)} \right) \right] \right] \nonumber \\
 &= \mathcal{S}_{(2, j)}.
\end{align}
Therefore, TFT of player $(1, j)$ is unbeatable against player $(2, j)$.
$\Box$

\subsection{Proof of Theorem \ref{thm:TFT_ZDS}}
\label{sec:proof_TFT_ZDS}
If the stage game is a nontrivial strongly payoff-monotonic game for player $(1, j)$, TFT of player $(1, j)$ satisfies
\begin{align}
 \sum_{a_{(1, j)}} r\left( a_{(1, j)} \right) T_{(1, j)} \left( a_{(1, j)} | \bm{a}^\prime \right) - r\left( a_{(1, j)}^\prime \right) &= r\left( a_{(2, j)}^\prime \right) - r\left( a_{(1, j)}^\prime \right) \nonumber \\
 &= s_j\left( \bm{a}_2^\prime \right) - s_j\left( \bm{a}_1^\prime \right).
\end{align}
By our assumption, $r$ is not a constant function.
Therefore, according to Definition \ref{def:fairZD}, TFT of player $(1, j)$ is a fair ZD strategy against player $(2, j)$ with $c_{a_{(1, j)}} =-r\left( a_{(1, j)} \right)$.
$\Box$

\subsection{Proof of Theorem \ref{thm:IIB_ZDS}}
\label{sec:proof_IIB_ZDS}
If the stage game is a nontrivial strongly payoff-monotonic game for player $(1, j)$, IIB of player $(1, j)$ satisfies
\begin{align}
 & \sum_{a_{(1, j)}} r\left( a_{(1, j)} \right) T_{(1, j)} \left( a_{(1, j)} | \bm{a}^\prime \right) - r\left( a_{(1, j)}^\prime \right) \nonumber \\
 &= r\left( a_{(2, j)}^\prime \right) \mathbb{I}\left( s_j\left( \bm{a}_1^\prime \right) < s_j\left( \bm{a}_2^\prime \right) \right) + r\left( a_{(1, j)}^\prime \right) \mathbb{I}\left( s_j\left( \bm{a}_1^\prime \right) \geq s_j\left( \bm{a}_2^\prime \right) \right) - r\left( a_{(1, j)}^\prime \right) \nonumber \\
 &= \left[ r\left( a_{(2, j)}^\prime \right) - r\left( a_{(1, j)}^\prime \right) \right] \mathbb{I}\left( s_j\left( \bm{a}_1^\prime \right) < s_j\left( \bm{a}_2^\prime \right) \right) \nonumber \\
 &= \left[ s_j\left( \bm{a}_2^\prime \right) - s_j\left( \bm{a}_1^\prime \right) \right] \mathbb{I}\left( s_j\left( \bm{a}_1^\prime \right) < s_j\left( \bm{a}_2^\prime \right) \right).
\end{align}
By our assumption, $r$ is not a constant function.
Therefore, according to Eq. (\ref{eq:ZDS}), IIB is a ZD strategy unilaterally enforcing Eq. (\ref{eq:linear_IIB}).
$\Box$

\subsection{Proof of Theorem \ref{thm:eIIB}}
\label{sec:proof_eIIB}
Suppose that the stage game is a game with an autocratic action for player $(1, j)$.
As far as $s_j\left( \bm{a}_1^\prime \right) \geq s_j\left( \bm{a}_2^\prime \right)$ holds, player $(1, j)$ continues to take $a_{(1, j)}^\prime$.
Once $s_j\left( \bm{a}_1^\prime \right) < s_j\left( \bm{a}_2^\prime \right)$ occurs, player $(1, j)$ using $\epsilon$-IIB chooses $a_j^{(\mathrm{H})}$ (in Definition \ref{def:weakPM}) with probability $\epsilon/|A_j|>0$.
It should be noted that the action $a_j^{(\mathrm{H})}$ is an absorbing state of $\epsilon$-IIB.
Therefore, the probability that player $(1, j)$ is beaten by player $(2, j)$ in a stage game $\tau$ times is at most $\left( 1- \epsilon/|A_j| \right)^{\tau-1}$.
Let $N$ be the total number of rounds that player $(1, j)$ is strictly beaten by the friend.
The above observation yields $N< \infty$ almost surely.
This implies
\begin{align}
 \liminf_{T\rightarrow \infty} \frac{1}{T}\sum_{t=1}^T \left[ s_j\left( \bm{a}_1^{(t)} \right) - s_j\left( \bm{a}_2^{(t)} \right) \right] \geq 0
\end{align}
almost surely.
Because of the finiteness of the payoff functions, if we take expectations in the above inequality and apply the Fatou's Lemma, we obtain $\mathcal{S}_{(1,j)} \geq \mathcal{S}_{(2,j)}$ almost surely.
This means that $\epsilon$-IIB of player $(1, j)$ is unbeatable against player $(2, j)$ almost surely.
$\Box$

\section*{Acknowledgement}
The author thanks the anonymous reviewers for their constructive comments.
This study was supported by Toyota Riken Scholar Program and JSPS KAKENHI Grant Number JP26K21335.

\appendix
\def\thesubsection{\Alph{section}.\arabic{subsection}}
\section{Stronger version of Theorems \ref{thm:IIB} and \ref{thm:TFT}}
\label{app:stronger_IIBTFT}
In the main text, we defined the unbeatability by Definition \ref{def:unbeatable}, and obtained Theorems \ref{thm:IIB} and \ref{thm:TFT}.
We can obtain stronger versions of Theorems \ref{thm:IIB} and \ref{thm:TFT} by defining unbeatability as in \cite{DOS2014}.
\begin{definition}
\label{def:es_unbeatable}
A memory-one strategy $T_{(1, j)}$ of player $(1, j)$ is \emph{essentially unbeatable} against player $(2, j)$ if the inequality
\begin{align}
 \sum_{t=1}^T \mathbb{E}\left[ s_j\left( \bm{a}_2^{(t)} \right) - s_j \left( \bm{a}_1^{(t)} \right) \right] &\leq \max_{\bm{a}}\left[ s_j\left( \bm{a}_2 \right) - s_j\left( \bm{a}_1 \right) \right] \quad (\forall T\geq 1)
\end{align}
holds regardless of behavior strategies of the other players and for arbitrary initial actions of player $(1, j)$.
\end{definition}
Because this definition does not assume the existence of $\mathcal{S}_{(\mu, j)}$, Definition \ref{def:es_unbeatable} is a stronger notion of unbeatability than Definition \ref{def:unbeatable}.
In fact, every essentially unbeatable strategy is unbeatable by considering the limit $\lim_{T\rightarrow \infty} \frac{1}{T}$ of the inequality.

For this definition, we obtain the following stronger theorems.
\begin{theorem}
\label{thm:IIB_es}
IIB of player $(1, j)$ is essentially unbeatable against player $(2, j)$ if and only if the stage game is a strongly payoff-monotonic game for player $(1, j)$.
\end{theorem}
\begin{proof}
(Necessity.)
Necessity is proved by the same method as that in the proof of Theorem \ref{thm:IIB}.

(Sufficiency.)
Suppose that the stage game is a strongly payoff-monotonic game for player $(1, j)$.
Because IIB of player $(1, j)$ takes $a_{(1, j)}^{(t)}$ at the $t$-th round with $t\geq 2$ such that
\begin{align}
 r\left( a_{(1, j)}^{(t)} \right) &= \max \left\{ r\left( a_{(1, j)}^{(1)} \right),  \max \left\{ r\left( a_{(2, j)}^{(t^\prime)} \right) | 1\leq t^\prime \leq t-1 \right\} \right\},
\end{align}
and because $A_j$ is a finite set, we obtain
\begin{align}
 \sum_{t=1}^T \left(s_j\left( \bm{a}_2^{(t)} \right) - s_j \left( \bm{a}_1^{(t)} \right) \right) &= \sum_{t=1}^T \left( r\left( a_{(2, j)}^{(t)} \right) - r\left( a_{(1, j)}^{(t)} \right) \right) \nonumber \\
 &= \sum_{t=1}^T \left( r\left( a_{(2, j)}^{(t)} \right) - \max \left\{ r\left( a_{(1, j)}^{(1)} \right),  \max \left\{ r\left( a_{(2, j)}^{(t^\prime)} \right) | 1\leq t^\prime \leq t-1 \right\} \right\} \right) \nonumber \\
 &\leq \sum_{t=1}^T r\left( a_{(2, j)}^{(t)} \right) - r\left( a_{(1, j)}^{(1)} \right) - \sum_{t=2}^T r\left( a_{(2, j)}^{(t-1)} \right) \nonumber \\
 &= r\left( a_{(2, j)}^{(T)} \right) - r\left( a_{(1, j)}^{(1)} \right) \nonumber \\
 &\leq \max_{a_j} r(a_j) - \min_{a_j} r(a_j) \nonumber \\
 &= \max_{\bm{a}_2} s_j\left( \bm{a}_2 \right) - \min_{\bm{a}_1} s_j\left( \bm{a}_1 \right).
\end{align}
Therefore, IIB of player $(1, j)$ is essentially unbeatable against player $(2, j)$.
\end{proof}

\begin{theorem}
\label{thm:TFT_es}
TFT of player $(1, j)$ is essentially unbeatable against player $(2, j)$ if and only if the stage game is a strongly payoff-monotonic game for player $(1, j)$.
\end{theorem}
\begin{proof}
(Necessity.)
Necessity is proved by the same method as that in the proof of Theorem \ref{thm:IIB}.

(Sufficiency.)
Suppose that the stage game is a strongly payoff-monotonic game for player $(1, j)$.
Because TFT of player $(1, j)$ takes
\begin{align}
 a_{(1, j)}^{(t)} &= a_{(2, j)}^{(t-1)}
\end{align}
at the $t$-th round with $t\geq 2$, and because $A_j$ is a finite set, we obtain
\begin{align}
 \sum_{t=1}^T \left(s_j\left( \bm{a}_2^{(t)} \right) - s_j \left( \bm{a}_1^{(t)} \right) \right) &= \sum_{t=1}^T \left( r\left( a_{(2, j)}^{(t)} \right) - r\left( a_{(1, j)}^{(t)} \right) \right) \nonumber \\
 &= \sum_{t=1}^T r\left( a_{(2, j)}^{(t)} \right) - r\left( a_{(1, j)}^{(1)} \right) - \sum_{t=2}^T r\left( a_{(2, j)}^{(t-1)} \right) \nonumber \\
 &= r\left( a_{(2, j)}^{(T)} \right) - r\left( a_{(1, j)}^{(1)} \right) \nonumber \\
 &\leq \max_{a_j} r(a_j) - \min_{a_j} r(a_j) \nonumber \\
 &= \max_{\bm{a}_2} s_j\left( \bm{a}_2 \right) - \min_{\bm{a}_1} s_j\left( \bm{a}_1 \right).
\end{align}
Therefore, TFT of player $(1, j)$ is essentially unbeatable against player $(2, j)$.
\end{proof}

\section{Numerical simulation in bandit-like game}
\label{app:bandit}
In this appendix, we investigate performance of imitation strategies in a bandit-like weakly payoff-monotonic game.
The payoff $s_1$ is given in Table \ref{table:bandit}.
In this game, $a_1^{(\mathrm{H})}=b_1$ and $a_1^{(\mathrm{L})}=b_{27}$ in Definition \ref{def:weakPM}.
Similarly as in the game in Table \ref{table:four-two}, if, for example, the action profile $a_{(1,1)}=b_{2}$, $a_{(1,2)}=d_3$, $a_{(2,1)}=b_{2}$, and $a_{(2,2)}=d_1$ is infinitely repeated, TFT and IIB of player $(1, 1)$ are beaten by a friend $(2, 1)$.
\begin{table}[tb]
  \centering
  \caption{Payoff $s_1$ in a bandit-like weakly payoff-monotonic game.}
  \begin{tabular}{|c|ccc|} \hline
    & $d_1$ & $d_2$ & $d_3$ \\ \hline
   $b_1$ & $1$ & $1$ & $1$ \\
   $b_2$ & $1$ & $1$ & $0$ \\
   $b_3$ & $1$ & $1$ & $-1$ \\
   $b_4$ & $1$ & $0$ & $1$ \\
   $b_5$ & $1$ & $0$ & $0$ \\
   $b_6$ & $1$ & $0$ & $-1$ \\
   $b_7$ & $1$ & $-1$ & $1$ \\
   $b_8$ & $1$ & $-1$ & $0$ \\
   $b_9$ & $1$ & $-1$ & $-1$ \\
   $b_{10}$ & $0$ & $1$ & $1$ \\
   $b_{11}$ & $0$ & $1$ & $0$ \\
   $b_{12}$ & $0$ & $1$ & $-1$ \\
   $b_{13}$ & $0$ & $0$ & $1$ \\
   $b_{14}$ & $0$ & $0$ & $0$ \\
   $b_{15}$ & $0$ & $0$ & $-1$ \\
   $b_{16}$ & $0$ & $-1$ & $1$ \\
   $b_{17}$ & $0$ & $-1$ & $0$ \\
   $b_{18}$ & $0$ & $-1$ & $-1$ \\
   $b_{19}$ & $-1$ & $1$ & $1$ \\
   $b_{20}$ & $-1$ & $1$ & $0$ \\
   $b_{21}$ & $-1$ & $1$ & $-1$ \\
   $b_{22}$ & $-1$ & $0$ & $1$ \\
   $b_{23}$ & $-1$ & $0$ & $0$ \\
   $b_{24}$ & $-1$ & $0$ & $-1$ \\
   $b_{25}$ & $-1$ & $-1$ & $1$ \\
   $b_{26}$ & $-1$ & $-1$ & $0$ \\
   $b_{27}$ & $-1$ & $-1$ & $-1$ \\  \hline
  \end{tabular}
  \label{table:bandit}
\end{table}

Here, we assume that both players $(1, 1)$ and $(2, 1)$ initially do not completely know the payoff function $s_1$ (Table \ref{table:bandit}).
We consider a situation where a friend $(2, 1)$ of player $(1,1)$ uses reinforcement learning \cite{SutBar2018}.
Concretely, the algorithm of reinforcement learning is given in Algorithm \ref{algo:RL}, where $Q(a_1)$ is called the action-value function.
Player $(1, 1)$ uses three imitation strategies: TFT, IIB, and $\epsilon$-IIB.
\begin{algorithm}
\caption{Reinforcement learning}
\label{algo:RL}
\begin{algorithmic}[1]
\REQUIRE Parameters $\eta \in (0, 1)$, $\alpha\in (0, 1]$, total time $T$
\ENSURE Actions $a_{(2,1)}^{(t)}$ for $1\leq t \leq T$
\STATE Set $Q\left( a_1 \right) = 0$ for all $a_1 \in A_1$
\FOR{$t=1$ to $T$}
\STATE Generate random variable $x\in [0, 1)$
\IF{$x<\eta$}
\STATE Choose $a_{(2,1)}^{(t)}$ from the uniform distribution on $A_1$
\ELSE
\STATE $a_{(2,1)}^{(t)} = \arg\max_{a_1} Q(a_1)$
\ENDIF
\STATE Update $Q\left( a_{(2,1)}^{(t)} \right) \leftarrow Q\left( a_{(2,1)}^{(t)} \right) + \alpha \left[ s_1\left( \bm{a}_2^{(t)} \right) - Q\left( a_{(2,1)}^{(t)} \right) \right]$
\ENDFOR
\end{algorithmic}
\end{algorithm}

\subsection{Random opponent}
\label{subsec:bandit_random}
\begin{figure}[tbp]
\begin{center}
\includegraphics[clip, width=8.0cm]{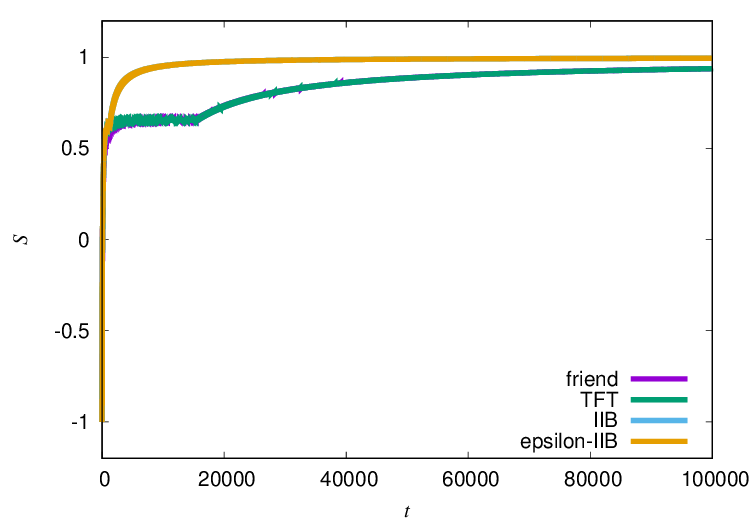}
\end{center}
\caption{The time-averaged payoffs of the focal player $(1, 1)$ and a friend $(2, 1)$ for the bandit-like weakly payoff-monotonic in Table \ref{table:bandit}. Players $(1,2)$ and $(2,2)$ take random actions in each round. Player $(2,1)$ adopts the reinforcement learning in Algorithm \ref{algo:RL} with $\eta=0.01$ and $\alpha=0.1$. Player $(1,1)$ adopts three imitation strategies: TFT (\ref{eq:TFT}), IIB (\ref{eq:IIB}), and $\epsilon$-IIB (\ref{eq:eIIB}) with $\epsilon=0.01$. The initial condition is $a_{(1,1)}=b_{27}$, $a_{(1,2)}=d_1$, $a_{(2,1)}=b_{27}$, and $a_{(2,2)}=d_1$. The time-averaged payoffs are calculated by one sample. The payoff of IIB almost overlaps with that of $\epsilon$-IIB.}
\label{fig:bandit}
\end{figure}

In Figure \ref{fig:bandit}, we display the numerical result for the case in which players $(1,2)$ and $(2,2)$ take random actions in each round.
The parameters are set to $\epsilon=0.01$, $\eta=0.01$, $\alpha=0.1$ and $T=10^5$.
The initial condition is $a_{(1,1)}=b_{27}$, $a_{(1,2)}=d_1$, $a_{(2,1)}=b_{27}$, and $a_{(2,2)}=d_1$.
The payoffs are calculated by the time-averaged payoffs of one sample.
In the figure, the payoff of IIB almost overlaps with that of $\epsilon$-IIB.

According to Theorem \ref{thm:eIIB}, $\epsilon$-IIB is unbeatable against the friend $(2, 1)$.
We can check that Figure \ref{fig:bandit} is consistent with this result.
Furthermore, TFT and IIB perform at least as well as the friend $(2, 1)$ in this simulation.
Therefore, in spite of Theorems \ref{thm:IIB} and \ref{thm:TFT}, TFT and IIB may practically be useful as strategies which cannot be beaten, if a friend is an excellent learner.
In addition, IIB and $\epsilon$-IIB outperform the friend.
This is because, once they find $a_1^{(\mathrm{H})}=b_1$, they repeat it forever.

\subsection{Adversarial opponent}
\begin{figure}[tbp]
\begin{center}
\includegraphics[clip, width=8.0cm]{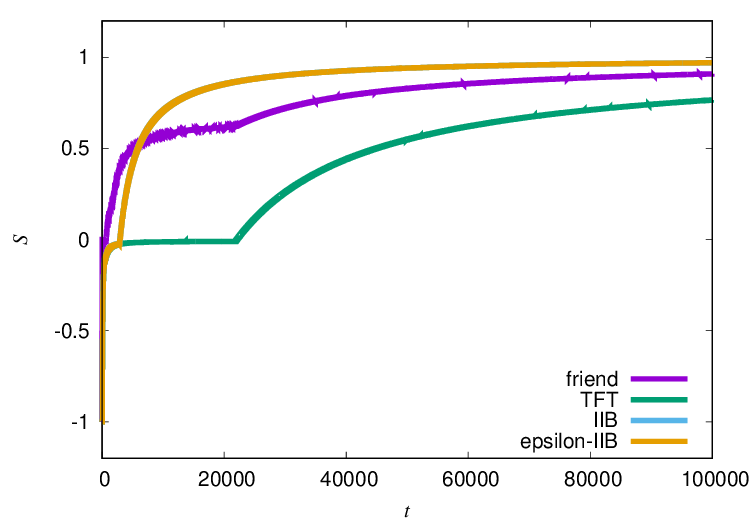}
\end{center}
\caption{The time-averaged payoffs of the focal player $(1, 1)$ and a friend $(2, 1)$ for the bandit-like weakly payoff-monotonic in Table \ref{table:bandit}. Player $(1,2)$ is a myopic adversarial opponent taking the action (\ref{eq:myopic_adversarial}) and player $(2,2)$ takes a random action in each round. Player $(2,1)$ adopts the reinforcement learning in Algorithm \ref{algo:RL} with $\eta=0.01$ and $\alpha=0.1$. Player $(1,1)$ adopts three imitation strategies: TFT (\ref{eq:TFT}), IIB (\ref{eq:IIB}), and $\epsilon$-IIB (\ref{eq:eIIB}) with $\epsilon=0.01$. The initial condition is $a_{(1,1)}=b_{27}$, $a_{(1,2)}=d_1$, $a_{(2,1)}=b_{27}$, and $a_{(2,2)}=d_1$. The time-averaged payoffs are calculated by one sample. The payoff of IIB almost overlaps with that of $\epsilon$-IIB.}
\label{fig:bandit_adversarial}
\end{figure}

In Figure \ref{fig:bandit_adversarial}, we display the numerical result for the case in which player $(1,2)$ is a myopic adversarial opponent taking
\begin{align}
 a_{(1,2)}^{(t)} &= \arg\min_{a_2} s_1\left( a_{(1,1)}^{(t)}, a_2 \right) \quad (\forall t\geq 1)
 \label{eq:myopic_adversarial}
\end{align}
and player $(2,2)$ takes a random action in each round.
The parameters are set to $\epsilon=0.01$, $\eta=0.01$, $\alpha=0.1$ and $T=10^5$.
The initial condition is $a_{(1,1)}=b_{27}$, $a_{(1,2)}=d_1$, $a_{(2,1)}=b_{27}$, and $a_{(2,2)}=d_1$.
The payoffs are calculated by the time-averaged payoffs of one sample.
In the figure, the payoff of IIB almost overlaps with that of $\epsilon$-IIB.

The result in Figure \ref{fig:bandit_adversarial} is almost the same as that in Appendix \ref{subsec:bandit_random}.
Even TFT seems to achieve equally good result with the friend after the friend $(2, 1)$ has learned the best action $b_1$.

\section*{References}

\bibliography{zds}

\end{document}